# Synchronous Multi-splitting Two-stage TOR Methods for Systems of Weakly Nonlinear Equations

Hwang Myong Gun,
Faculty of Mathematics, **Kim Il Sung** University

**Abstract.** For the large sparse systems of weakly nonlinear equations arising in the discretizations of many classical differential and integral equations, this paper presents a class of synchronous parallel multi-splitting two-stage two-parameter over-relaxation (TOR) methods for getting their solutions by the high-speed multiprocessor systems. Under suitable assumptions, we study the global convergence properties of these synchronous multi-splitting two-stage TOR methods.

## 1 Introduction

Consider the system of weakly nonlinear equations

$$Ax = G(x), A \in R^{n \times n}, G: R^n \to R^n. \qquad (1)$$

Where the matrix $A \in R^{n \times n}$ is large sparse, and usually, monotones; the nonlinear mapping $G: R^n \to R^n$ is bounded, but its derivative may not exist in the ordinary meaning. Since this class of weakly nonlinear systems has bounteous applicable backgrounds and special mathematical structure, Bai [1] has studied in depth the serial and parallel two-stage iterative methods for approximately getting its solution, and described in detail the convergence properties of these methods.

Bai and Wang [2] will further discuss the local and the global convergence properties of the parallel multi-splitting two-stage iterative methods under different assumptions from those in [1], in the sense of positive definite ordering. In finding the solution of the special type of the practical problem, numerical solving method with the general purpose is not always effective, so it is important to study the solving methods that reflect the concrete structure of the given problem. In [4] it was considered the relation between the separation of matrix and the two-stage iterative method.

In accordance with the concrete characterization of the multiprocessor system and the large sparse property of the system of weakly nonlinear equations (1), by the two-parameter over-relaxation technique of the linear iterative method, we first present a multi-splitting two-stage TOR method, which particularly uses the known TOR-like iteration ([3]) as inner iterations and is a relaxed variant of the afore-presented iterative methods. Then we thoroughly set up the global convergence theories of these now methods when the system matrix $A \in R^{n \times n}$ is either a $H$-matrix or a monotone matrix, and the mapping $G: R^n \to R^n$ is a bounded mapping.

## 2 Synchronous multi-splitting two-stage TOR method

For a given positive integer $p$, let $A = B_l - C_l (l = 1, 2, \cdots, p)$ be $p$ splitting of the matrix $A \in R^{n \times n}$, i.e. $\det(B_l) \neq 0 \, (l = 1, 2, \cdots, p)$; $B_l = M_l - N_l (l = 1, 2, \cdots, p)$ be splittings of the matrices $B_l \in R^{n \times n}$ $(l = 1, 2, \cdots, p)$, respectively, and $E_l \in R^{n \times n}$ be nonnegative diagonal matrices satisfying

$$\sum_{l=1}^{p} E_l = I \text{ (the } n \times n \text{ identity matrix)}.$$

Then the collection of triples $(B_l, C_l, E_l)(l = 1, 2, \cdots, p)$ is called a multi-splitting of the matrix A; and the collection of quintuples $(B_l : M_l, N_l; C_l; E_l)(l = 1, 2, \cdots, p)$ is called a two-stage multi-splitting of the matrix A. In the sequel, we assume that the considered multiprocessor system is made up of $p$ processors, referred to $\Pr oc(1), \Pr oc(2), \cdots, \Pr oc(p)$; and without loss of generality, we stipulate that the host processor, say $\Pr oc(0)$), may be undertaken by any of the $p$ processors.

For each $l = 1, 2, \cdots, p$, let $D_l = diag(B_l)$, $V_l + V_l^* \in R^{n \times n}$ be strictly lower triangular matrices of the matrices $(-B_l)$, respectively, and $U_l \in R^{n \times n}$ be zero-diagonal matrices such that $B_l = D_l - (V_l + V_l^*) - U_l$. Then we naturally get a special two-stage multi-splitting of the matrix A, denoted as

$$(B_l : D_l + (V_l + V_l^*), U_l; C_l; E_l)(l = 1, 2, \cdots, p).$$

In addition, for some real constants $\alpha$ and $\beta$ ($\alpha, \beta \geq 0, \alpha + \beta \neq 0$), define

$$M_l(\alpha, \beta) = (2D - \alpha V_l - \beta V_l^*)/(\alpha + \beta)$$
$$N_l(\alpha, \beta) = \{[2 - (\alpha + \beta)]D + (\alpha + \beta)U_l - \alpha V_l^* - \beta V_l\}/(\alpha + \beta)$$

Then a multi-splitting two-stage TOR method for parallel solving of the system of weakly nonlinear equation (1) can be detrained.

**Algorithm** (Synchronous multi-splitting TOR method)

Given an initial vector $x^0 \in R^n$.
FOR $i = 0, 1, 2, \cdots$ until $\{x^p\}$ convergence DO
Begin
    CoBegin $\Pr oc(1), \Pr oc(2), \cdots, \Pr oc(p)$
        $\Pr oc(l)$ $(l = 1, 2, \cdots, n)$:
            $x^{i, l, 0} = x^i$
            FOR $k = 0$ TO $s_l(i) - 1$ DO
$$x^{i, l, k+1} = (2D_l - \alpha V_l - \beta V_l^*)^{-1}\{[2 - (\alpha + \beta) + (\alpha + \beta)U_l - \alpha V_l^* - \beta V_l\}x^{i, l, k}$$
$$+ [(\alpha + \beta)(C_l x^i + G(x^i))]$$
    CoEnd
    $\Pr oc(0) : x^{i+1} = \sum_{i=1}^{p} E_l x^{i+1, l}$
End

Here $\{s_l(i)\}_{i=0}^{\infty}$ $(l = 1, 2, \cdots, p)$ are positive integer sequences.
This algorithm can be expressed in the matrix-vector form:

$$x^{i+1} = \sum_{l=1}^{p} E_l\{[M_l(\alpha, \beta)^{-1}N_l(\alpha, \beta)]^{s_l(i)}x^i + \sum_{j=0}^{s_l(i)-1}[M_l(\alpha, \beta)^{-1}N_l(\alpha, \beta)]^j \quad (2)$$
$$M_l(\alpha, \beta)^{-1}(C_l x^i + G(x^i)]$$

Note that iterative (2) includes two arbitrary parameters $\alpha$ and $\beta$, their suitable adjustment can greatly improve the convergence property of this method. Moreover, iteration (2) be
  synchronous multi-splitting two-stage AOR method, when
  $\alpha = 2r, \alpha + \beta = 2w$,

synchronous multi-splitting two-stage SOR method, when $\alpha = 2w, \beta = 0$,
synchronous multi-splitting two-stage GS method, when $\alpha = 2, \beta = 0$,
synchronous multi-splitting two-stage JOR method, when $\alpha = 0, \beta = 2w$,
synchronous multi-splitting two-stage Jacobi iteration method, when $\alpha = 0, \beta = 2$.

In fact, iteration (2) is nonlinear extensions of the matrix multi-splitting TOR methods for the system of linear equations (see [ 6 ]), and is parallel variants of the serial two-stage TOR methods for system of weakly nonlinear equations (see [1]).

Let $\vec{\alpha} = (\alpha_1, \alpha_2, \cdots, \alpha_p)^T$, $\vec{\beta} = (\beta_1, \beta_2, \cdots, \beta_p)^T$. Then we get the multiple synchronous multi-splitting TOR method:

$$x^{i+1} = \sum_{l=1}^{p} E_l \{[M_l(\vec{\alpha}, \vec{\beta})^{-1} N_l(\vec{\alpha}, \vec{\beta})]^{s_l(i)} x^i + \sum_{j=0}^{s_l(i)-1} [M_l(\vec{\alpha}, \vec{\beta})^{-1} N_l(\vec{\alpha}, \vec{\beta})]^j \quad (3)$$
$$M_l(\vec{\alpha}, \vec{\beta})^{-1} (C_l x^i + G(x^i)], \quad i = 0, 1, 2, \cdots.$$

## 3 Global convergence theories

In this section, we will discuss the global convergence properties of the multi-splitting two-stage TOR methods. For this purpose, we will closely follow the notations and concepts introduced in [1] in the sequel. In particular, we denote by $|\cdot|$ the absolute value of either a vector or a matrix, and $<\cdot>$ the comparison matrix of the corresponding matrix. We call a mapping $G: R^n \to R^n$ $P$ – bounded if there exists a nonnegative matrix $P \in R^{n \times n}$ such that

$$|G(x) - G(y)| \leq P |x - y| \quad \forall x, y \in R^n.$$

For the existence and uniqueness of the solution of the system of weakly nonlinear equations (1), Bai [1] has proved the following theorem.

**Theorem 1** (see [1]) Let $A \in R^{n \times n}$ be a nonsingular matrix, and $G: R^n \to R^n$ be $P$ – bounded. Then the system of weakly nonlinear equations (1) has a unique solution $x^* \in R^n$ provided either of the following two conditions is satisfied:

(a) $A \in R^{n \times n}$ Is a monotone matrix and $\rho(A^{-1} P) < 1$.

(b) $A \in R^{n \times n}$ Is an $H$ – matrix and $\rho(<A^{-1}>P)<1$.

Now, we demonstrate the global convergence of new method when the system matrix $A \in R^{n \times n}$ is an $H$ – matrix and nonlinear mapping $G: R^n \to R^n$ is a $P$ – bounded mapping.

**Theorem 2** Let $A \in R^{n \times n}$ be an $H$ – matrix with $D = diag(A)$ and $A = D - B$, $(B_l, C_l, E_l)$ $(l = 1, 2, \cdots, p)$ be its multi-splitting satisfying $<A> = <B_l> - |C_l|$. Assume $G: R^n \to R^n$ to be a $P$ – bounded mapping and $\rho(<A^{-1}>P) < 1$. For any starting vector $x^0 \in R^n$ and any number sequences $\{s_l(i)\}_{i=0}^{\infty}$ $(l = 1, 2, \cdots, p)$ of the inner iterations satisfying $s_l(i) > 1, (l = 1, 2, \cdots, p)$, if

$$(B_l : D_l - (V_l + V_l^*), U_l; C_l; E_l) \ (l = 1, 2, \cdots, p)$$

is a two-stage multi-splitting of the matrix $A \in R^{n \times n}$ and

$$D_l \equiv D, <B_l> = |D_l| - |V_l| - |V_l^*| - |U_l|, l = 1, 2, \cdots, p,$$

Then the sequences $\{x^i\}$ generated by (2) and (3) converges to the unique solution $x^* \in R^n$ of the system of weakly nonlinear equations (1) provided the parameters $\alpha$ and $\beta$ satisfy

$$0 < \alpha + \beta < \frac{4}{1 + \rho(|D|^{-1}(|B|+P))}$$

And

$$0 < \alpha_l + \beta_l < \frac{4}{1 + \rho(|D|^{-1}(|B|+P))},$$

respectively.

*Proof.* We first demonstrate for iteration (2). By theorem 1 we know that there exists a unique vector $x^* \in R^n$ such that $Ax^* = G(x^*)$. Since iteration (2) is consistent with the weakly nonlinear equations (1), corresponding to (2) $x^*$ also satisfies

$$x^* = \sum_{l=1}^{p} E_l\{[M_l(\alpha,\beta)^{-1} N_l(\alpha,\beta)]^{s_l(i)} x^* + \sum_{j=0}^{s_l(i)-1} [M_l(\alpha,\beta)^{-1} N_l(\alpha,\beta)]^j M_l(\alpha,\beta)^{-1} \quad (4)$$
$$(C_l x^* + G(x^*)], \quad i = 0, 1, 2, \cdots.$$

Now, if we introduce error vectors $\varepsilon^i = x^i - x^* (i = 0, 1, 2, \cdots)$, by subtracting (4) from (2) we can easily get the recursive formula :

$$\varepsilon^{i+1} = \sum_{l=1}^{p} E_l\{[M_l(\alpha,\beta)^{-1} N_l(\alpha,\beta)]^{s_l(i)} \varepsilon^i + \sum_{j=0}^{s_l(i)-1} [M_l(\alpha,\beta)^{-1} N_l(\alpha,\beta)]^j M_l(\alpha,\beta)^{-1} \quad (5)$$
$$(C_l \varepsilon^i + G(x^i) - G(x^*)], \quad i = 0, 1, 2, \cdots.$$

Noticing that $A \in R^{n \times n}$ is an $H$–matrix, we know that the matrices and are all $H$–matrices. Hence, there hold

$$|B_l| \leq\!\!< B_l >^{-1}, |M_l(\alpha,\beta)^{-1}| \leq\!\!< M_l(\alpha,\beta) >^{-1}, l = 1, 2, \cdots, p. \quad (6)$$

Noticing that for $l = 1, 2, \cdots, p$,

$$|M_l(\alpha,\beta)^{-1}| = |(\alpha+\beta)(2D - \alpha V_l - \beta V_l^*)^{-1}|$$
$$\leq (\alpha+\beta)(2|D| - \alpha|V|_l - \beta|V_l^*|)^{-1} \equiv \hat{M}_l(\alpha,\beta)^{-1}$$
$$|N_l(\alpha,\beta)| = |\{[2-(\alpha+\beta)]D + (\alpha+\beta)U_l - \alpha V_l^* - \beta V_l\}/(\alpha+\beta)|$$
$$\leq \{|2-(\alpha+\beta)||D| + (\alpha+\beta)|U_l| - \alpha|V_l^*| - \beta|V_l|\}/(\alpha+\beta),$$
$$\equiv \hat{N}_l(\alpha,\beta)$$

By making use of the $P$–bounded property of the mapping $G: R^n \to R^n$ we obtain from (5)

$$|\varepsilon^{i+1}| = |\sum_{l=1}^{p} E_l[M_l(\alpha,\beta)^{-1} N_l(\alpha,\beta)]^{s_l(i)} \varepsilon^i + \sum_{j=0}^{s_l(i)-1} [M_l(\alpha,\beta)^{-1} N_l(\alpha,\beta)]^j$$
$$M_l(\alpha,\beta)^{-1}(C_l \varepsilon^i + G(x^i) - G(x^*)]$$
$$\leq \sum_{l=1}^{p} E_l[|M_l(\alpha,\beta)^{-1} N_l(\alpha,\beta)|^{s_l(i)}|\varepsilon^i| + \sum_{j=0}^{s_l(i)-1} |M_l(\alpha,\beta)^{-1} N_l(\alpha,\beta)|^j$$
$$|M_l(\alpha,\beta)^{-1}|(|C_l||\varepsilon^i| + |G(x^i) - G(x^*)|)]$$
$$\leq \sum_{l=1}^{p} E_l[(\hat{M}_l(\alpha,\beta)^{-1} \hat{N}_l(\alpha,\beta))^{s_l(i)} + \sum_{j=0}^{s_l(i)-1} (\hat{M}_l(\alpha,\beta)^{-1} \hat{N}_l(\alpha,\beta))^j$$
$$\hat{M}_l(\alpha,\beta)^{-1}(|C_l| + P)]|\varepsilon^i|$$

$$\equiv \hat{T}_i(\alpha, \beta)|\varepsilon^i| \leq \hat{T}_i(\alpha, \beta)\hat{T}_{i-1}(\alpha, \beta)\cdots \hat{T}_0(\alpha, \beta)|\varepsilon^0|. \tag{7}$$

Where we have used the notations

$$\hat{T}_i(\alpha, \beta) = \sum_{l=1}^{p} E_l[(\hat{M}_l(\alpha, \beta)^{-1}\hat{N}_l(\alpha, \beta))^{s_l(i)} + \sum_{j=0}^{s_l(i)-1}(\hat{M}_l(\alpha, \beta)^{-1}\hat{N}_l(\alpha, \beta))^j \\ \hat{M}_l(\alpha, \beta)^{-1}(|C_l|+P)]. \tag{8}$$

Define matrices

$$\hat{A}(\alpha+\beta) = \frac{2-(\alpha+\beta)-|2-(\alpha+\beta)|}{\alpha+\beta}|D| + <A> - P,$$

$$\hat{B}_l(\alpha+\beta) = \frac{2-(\alpha+\beta)-|2-(\alpha+\beta)|}{\alpha+\beta}|D| + <B_l>,$$

$$\hat{C}_l(\alpha+\beta) = |C_l| + P.$$

Then, there hold

$$\hat{A}(\alpha+\beta) = \hat{B}_l(\alpha+\beta) - \hat{C}_l(\alpha+\beta),$$
$$\hat{B}_l(\alpha+\beta) = \hat{M}_l(\alpha, \beta) - \hat{N}_l(\alpha, \beta),$$
$$l = 1, 2, \cdots, p.$$

Clearly, $\hat{C}_l(\alpha+\beta) \geq 0 (l = 1, 2, \cdots, p)$, and we know that

$$\hat{B}_l(\alpha+\beta) = \hat{M}_l(\alpha, \beta) - \hat{N}_l(\alpha, \beta), l = 1, 2, \cdots, p.$$

are M-splitting. So, to fulfill this proof we only need to verify that $\hat{A}(\alpha+\beta)$ and $\hat{B}_l(\alpha+\beta)$ $(l = 1, 2, \cdots, p)$ are monotone matrices under the conditions. Noticing

$$\hat{A}(\alpha+\beta) = \hat{B}_l(\alpha+\beta) - (|C_l|+P) \leq \hat{B}_l(\alpha+\beta) \leq <B_l> \leq <D>, l = 1, 2, \cdots, p,$$

we hence only need to test that $\hat{A}(\alpha+\beta)$ is a monotone matrix.

In fact, let $|Q| = |B| + P$, $\hat{R} = |D| - |Q|$. Since

$$\hat{R} = |D| - |Q| = |D| - |B| - P = <A> - P,$$

and $\rho(<A^{-1}>P) < 1$, we easily see that $\hat{R}$ is a monotone matrix. Hence, $\rho(|D|^{-1}|Q|) < 1$. Considering

$$\hat{A}(\alpha+\beta) = \frac{2-|2-(\alpha+\beta)|}{\alpha+\beta}|D| - |Q|,$$

we immediately know that $\hat{A}(\alpha+\beta)$ is a monotone matrix when

$$0 < \alpha+\beta < \frac{4}{1+\rho(|D|^{-1}(|B|+P))}.$$

Because $\hat{A}(\alpha+\beta)$ and $\hat{B}_l(\alpha+\beta)$ $(l=1,2,\cdots,p)$ are monotone matrices and

$$\hat{B}_l(\alpha+\beta) - \hat{A}(\alpha+\beta) = |C_l| + P = <B_l> - (<A> + P),$$

We can obtain

$$\begin{aligned}
\hat{T}_i(\alpha,\beta) &= \sum_{l=1}^{p} E_l [(\hat{M}_l(\alpha,\beta)^{-1}\hat{N}_l(\alpha,\beta))^{s_l(i)} + \sum_{j=0}^{s_l(i)-1}(\hat{M}_l(\alpha,\beta)^{-1}\hat{N}_l(\alpha,\beta))^j \\
&\quad \hat{M}_l(\alpha,\beta)^{-1}(|C_l|+P)] \\
&= \sum_{l=1}^{p} E_l [(\hat{M}_l(\alpha,\beta)^{-1}\hat{N}_l(\alpha,\beta))^{s_l(i)} + \sum_{j=0}^{s_l(i)-1}(\hat{M}_l(\alpha,\beta)^{-1}\hat{N}_l(\alpha,\beta))^j \\
&\quad \hat{M}_l(\alpha,\beta)^{-1}<B_l> - \sum_{j=0}^{s_l(i)-1}(\hat{M}_l(\alpha,\beta)^{-1}\hat{N}_l(\alpha,\beta))^j \hat{M}_l(\alpha,\beta)^{-1}(<A>-P)] \\
&= I - \sum_{l=1}^{p} E_l \sum_{j=0}^{s_l(i)-1}(\hat{M}_l(\alpha,\beta)^{-1}\hat{N}_l(\alpha,\beta))^j \hat{M}_l(\alpha,\beta)^{-1}(<A>-P).
\end{aligned} \quad (9)$$

Therefore, there exists a positive vector $u \in R^n$ such that $(<A>-P)u = e$, where $e = (1,1,\cdots,1)^T \in R^n$. We obtain from (9) that

$$\hat{T}_i(\alpha,\beta)u = u - \sum_{l=1}^{p} E_l \sum_{j=0}^{s_l(i)-1}(\hat{M}_l(\alpha,\beta)^{-1}\hat{N}_l(\alpha,\beta))^j \hat{M}_l(\alpha,\beta)^{-1} e$$

$$\leq u - \sum_{l=1}^{p} E_l \hat{M}_l(\alpha, \beta)^{-1} e$$
$$\leq u - \sum_{l=1}^{p} E_l (1-\theta) u = \theta u.$$

Take $\sigma > 0$ such that $|\varepsilon^0| \leq \sigma u$. Then by (7) we immediately have

$$|\varepsilon^{i+1}| \leq \hat{T}_i(\alpha, \beta) \hat{T}_{i-1}(\alpha, \beta) \cdots \hat{T}_0(\alpha, \beta) \sigma u \leq \sigma \theta^{i+1} u \to 0.$$

This obviously implies $\lim_{i \to \infty} \varepsilon^i = 0$, or in other words, $\lim_{i \to \infty} x^i = x^*$. ∎

The proof for iteration (3) is analogous to that of iteration (2), so it is omitted here. More specifically, when the system matrix $A \in R^{n \times n}$ is a monotone matrix, theorem 2 directly results in the following global convergence theorem for the new multi-splitting two-stage TOR methods.

**Theorem 3.** Let $A \in R^{n \times n}$ be a monotone matrix with $D = diag(A)$, $A = D - B$, and $(B_l, C_l, E_l)(l = 1, 2, \cdots, p)$ be its multi-splitting with $(B_l, C_l, E_l)(l = 1, 2, \cdots, p)$ being regular splitting. Assume $G: R^n \to R^n$ to be a $P$–bounded mapping and $\rho(A^{-1}P) < 1$. For any starting vector $x^0 \in R^n$ and any number sequences $\{s_l(i)\}_{i=0}^{\infty} (l = 1, 2, \cdots, p)$ of the inner iterations satisfying $s_l(i) > 1, (l = 1, 2, \cdots, p)$, if $(B_l : D_l - (V_l + V_l^*), U_l; C_l; E_l)$ $(l = 1, 2, \cdots, p)$ is a two-stage multi-splitting of the matrix $A \in R^{n \times n}$ with $D_l \geq 0, V_l + V_l^* \geq 0, U_l \geq 0, l = 1, 2, \cdots, p$ and , then the sequences $\{x^i\}$ generated by (2) and (3) converges to the unique solution $x^* \in R^n$ of the system of weakly nonlinear equations (1) provided the parameters $\alpha$ and $\beta$ satisfy $0 < \alpha + \beta < 2$ and $0 < \alpha_l + \beta_l < 2$, respectively.